\documentclass{rmf-d}
\usepackage{nopageno,rmfbib,multicol,times,epsf,amsmath,amssymb,cite}
\usepackage[latin1]{inputenc}
\usepackage[]{caption2}
\usepackage{graphics}
\usepackage{float}
\usepackage{graphicx}
\usepackage{color}
\usepackage[mathscr,scaled=1.15]{urwchancal}
\DeclareFontFamily{OT1}{pzc}{}
\DeclareFontShape{OT1}{pzc}{m}{it}%
{<-> s * [1.15] pzcmi7t}{}
\DeclareMathAlphabet{\mathpzc}{OT1}{pzc}{m}{it}

\def\rmfcornisa{Research OR Education of Physics \hfill\rmf\ 
\hfill MES A\~NO}

\def\rmfcintilla{{\it Rev.\ Mex.\ Fis.\ S.\/} }
\clearpage \rmfcaptionstyle 
\begin{document}
\title{   Highlights of pion and kaon structure from continuum analyses
\vspace{-6pt}}
\author{ Kh\'epani Raya$^{1,2}$,\;\;Jos\'e Rodr\'iguez-Quintero$^2$ }
\address{ $^1$Departamento de F\'isica Te\'orica y del Cosmos, Universidad de Granada, E-18071, Granada, Spain \\
$^2$Department of Integrated Sciences and Center for Advanced Studies in Physics,
Mathematics and Computation; University of Huelva, E-21071 Huelva; Spain. }
\author{ }
\address{ }
\author{ }
\address{ }
\author{ }
\address{ }
\author{ }
\address{ }
\maketitle
\recibido{day month year}{day month year
\vspace{-12pt}}
\begin{abstract}
\vspace{1em} One of the biggest challenges in contemporary physics is understanding the origin and dynamics of the internal structure of hadrons which, at a fundamental level, is described by quantum chromodynamics (QCD). Taking great prominence amongst hadrons are pions and kaons which, despite being the lightest hadrons in nature, their very existence is intimately connected to those mechanisms responsible for almost all of the mass of the visible matter. In this manuscript we discuss many aspects of the pion and kaon structure via light front wave functions and generalized parton distributions, and a collection of other distributions and structural properties that are inferred therefrom.  \vspace{1em}
\end{abstract}
\keys{Nambu-Goldstone bosons, Generalized parton distributions, Continuum Schwinger Methods\vspace{-4pt}}
\begin{multicols}{2}

\section{Introduction}
Quantum chromodynamics (QCD) is the theory of strong interactions in the Standard Model. It is supposed to explain the existence and properties of hadrons in terms of fundamental degrees of freedom, quarks and gluons, and their interactions. In very specific kinematic domains, usually involving high energies, perturbative QCD rigorously connects hadronic observables with the first principles of the theory~\cite{Lepage:1980fj,Lepage:1979zb,Berger:1979du,Farrar:1975yb}. Nonetheless, in the daily life domain, emergent phenomena arise for reasons that turn out to be not very obvious~\cite{Roberts:2020hiw}. One one hand, confinement entails that quarks and gluons do not appear isolated in nature; instead, they are the building blocks of composite colorless particles dubbed hadrons. On the other hand, the non perturbative nature of QCD triggers the dynamical generation of mass, and thus the emergence of hadron masses (EHM)~\footnote{In fact, only about $2\%$ of the mass in the visible universe can be attibuted to the Higgs mechanism alone.}. Confinement and EHM are naturally connected with QCD's strong running coupling. Our modern understanding indicates that, in the infrarred, the running coupling in QCD is enhanced and saturates at a finite value and, as we move away from this domain, the coupling smoothly acquires its perturbative profile~\cite{Rodriguez-Quintero:2019fyc,Cui:2019dwv,Rodriguez-Quintero:2018wma,Binosi:2016nme}.

Being the most abundant and stable hadron in Nature, the proton has played a crucial role in improving our understanding of QCD and hadron physics~\cite{Hofstadter:1956qs, Breidenbach:1969kd,Ellis:1996mzs}, then it is an ideal platform to inquire about the emergent phenomena in QCD: for instance, one must be capable to explain the origin of the massiveness of the proton ($m_P \sim 1$ GeV) and its size ($r_P \sim 1$ fm)~\cite{Roberts:2020hiw,Aznauryan:2012ba}.  Notwithstanding, studying the lightest hadrons in nature, pions and kaons, must be in physicists minds as well: being the Nambu-Goldstone (NG) bosons of dynamical chiral symmetry breaking (DCSB), their mere existance owes to the same underlying mechanisms of dynamical generation of mass in QCD~\cite{Roberts:2021nhw,Arrington:2021biu}. Thus, massiveness of the proton is intimately connected with the existence and \emph{masslessness} of NG bosons. Furthermore, in the absence of the Higgs mechanism, both pions and kaons would be massless and indistinguishable from each other; then, the interplay between QCD and Higgs mass generation triggers their structural differences. 

On the eve of next generation experiments with meson targets~\cite{Arrington:2021biu,Aguilar:2019teb,Anderle:2021wcy,Adams:2018pwt}, which would expose an array of structural properties of pions and kaons, robust theoretical predictions and explanations must be readily available. Our present analysis is based upon Ref.~\cite{Raya:2021zrz}.
We discuss many aspects of $\pi-K$ structure and its connection with confinement and the EHM, as inferred from generalized parton distributions (GPDs), obtained via light-front wavefunctions (LFWFs). The manuscript is organized as follows: Section 2 introduces some basics about LFWFs and its connection with the distribution amplitudes (PDAs) and distribution functions (PDFs); how predictions for PDAs and PDFs can be employed to inform the LFWF are also addressed.  Generalized parton distributions (GPDs) are introduced in Section 3, exposing its connection with electromagnetic and gravitational form factors, as well as an array of distributions that can be derived therefrom. In Section 4 we discuss some algebraic insights that can be obtained from factorized models for LFWFs. Conclusions and scope are presented in Section 5.

\section{Meson light-front wavefunctions}
The leading twist meson LFWF is obtained from the projection of the Poincar\'e covariant Bethe-Salpeter wavefunction (BSWF), $\chi_\textbf{P}$, onto the light front. For a $u$-quark in a pseudoscalar  meson $\textbf{P}=u\bar{h}$, it might be written ($k_-:=k-P_{\textbf{P}}/2$):
\begin{equation}
\label{eq:defLFWF}
\psi_{\textbf{P}u}^{\uparrow \downarrow}(x,k_\perp^2;\zeta_H)=Z_2 \text{tr}_{\text{CD}}  \int \frac{dk_\parallel}{\pi}   \delta^x_n(k) \gamma_5 \gamma \cdot n \chi_\textbf{P}(k_-;P_{\textbf{P}})\;,
\end{equation}
where $\text{tr}_{\text{CD}}$ indicates the trace over color and spinor indices, and the arrows $\uparrow \downarrow$ specify that we are restraining ourselves to the helicity-0 component of the LFWF. The pieces enterig Eq.~\eqref{eq:defLFWF} are defined as usual: $x$ is the light-front momentum fraction of the meson's total momentum $P_{\textbf{P}}$, and $P^2_{\textbf{P}}=-m_{\textbf{P}}^2$, with $m_{\textbf{P}}$ the meson mass (when obvious, we shall avoid the labeling ${\textbf{P}}$); $n$ is light-like four vector, namely $n^2=0$ and $n \cdot P = -m_{\textbf{P}}$ in the meson rest-frame; $k=k_\parallel+k_\perp$ such that $k_\parallel=(k^+,k^-,\textbf{0})$ in light-cone coordinates; $\delta^x_{n}(\#) = \delta( n \cdot \#  - x n \cdot P_\textbf{P})$ and $Z_2$ is simply the quark wavefunction renormalization constant. Note the dependence on $\zeta_H$, the hadronic scale, which indicates that the LFWF represents the hadron on a scale in which the fully dressed valence quarks express all the properties of the hadron; in particular, all the hadron momentum is contained within the fully dressed valence quarks (thus nothing in the gluon and sea quarks).

The so called leading-twist PDA is obtained by integrating the $k_\perp$ degrees of freedom of $\psi_{\textbf{P}u}^{\uparrow\downarrow}$, namely: 
\begin{equation}
\label{eq:PDALFWF}
f_{\textbf{P}} \varphi^{u}_{\textbf{P}}(x;\zeta_H) =  \int \frac{d^2k_\perp}{16 \pi^3} \psi^{\uparrow \downarrow}_{\textbf{P}u}\left(x,k_\perp^2;\zeta_H \right)\;, 
\end{equation}
Herein, $\varphi^{u}_\textbf{P}$ denotes the $u$-in-$\textbf{P}$ PDA and $f_{\textbf{P}}$ is the meson leptonic decay constant, which is an order parameter of DCSB. Written in this way, $\varphi^{u}_\textbf{P}$ is unit normalized and, owing to momentum conservation, the antiquark PDA is simply
\begin{equation}
    \label{eq:PDAanti}
    \varphi^{\bar{h}}_\textbf{P}(x;\zeta)=\varphi^{u}_\textbf{P}(1-x;\zeta)\;;
\end{equation}
the above relation being valid for any scale $\zeta$, even after evolution (ERBL)~\cite{Lepage:1979zb,Efremov:1979qk,Lepage:1980fj}. On the other hand, the valence quark and antiquark PDFs, as derived from the LFWF, read  

\begin{subequations}
\label{eq:uMzetaH}
\begin{align}
u^{\textbf{P}}(x;\zeta_H) &= \int \frac{d^2k_\perp}{16\pi^3} \left| \psi^{\uparrow \downarrow}_{\textbf{P}u}\left(x,k_\perp^2;\zeta_H \right) \right|^2 \,, \label{eq:LFWFPDF} \\
\overline{h}^{\textbf{P}}(x;\zeta_H)&=u^{\textbf{P}}(1-x;\zeta_H)\,.
\end{align}
\end{subequations}
Unlike the PDA, this simple connection between the quark and antiquark PDFs is only valid at $\zeta_H$. The reason is that the PDF obeys a different set of evolution equations (DGLAP)~\cite{Dokshitzer:1977sg, Gribov:1972ri, Lipatov:1974qm, Altarelli:1977zs}, which explicitly incorporate sea-quark and gluons degrees of freedom.

It is clear that Eqs.~(\ref{eq:PDALFWF}, \ref{eq:LFWFPDF}) expose a bridge between PDA and PDF, via LFWF; a sensible representation for the latter is thus required. For this purpose, it is worth reminding that the BSWF is defined as the sandwich of the Bethe-Salpeter amplitude (BSA) and quark propagators, namely:
\begin{equation}
    \label{eq:defBSA}
    \chi_\textbf{P}(k_-;P)=S_u(k) \Gamma_\textbf{P}(k_-;P)S_{\bar{h}}(k-P)\;;
\end{equation}
where the pseudoscalar meson BSA, $\Gamma_{\textbf{P}}$, is decomposed into 4 Dirac structures,
    \begin{eqnarray}
\label{eq:BSAstruct}
    \Gamma_{\textbf{P}}(q;P) &=& \gamma_5[i \mathbb{E}_{\textbf{P}}(q;P)+\gamma \cdot P \mathbb{F}_{\textbf{P}}(q;P) \\
    &+& \gamma \cdot q \mathbb{G}_{\textbf{P}}(q;P) + q_\mu \sigma_{\mu \nu} P_\nu \mathbb{H}_{\textbf{P}}(q;P)]\;,\nonumber
\end{eqnarray}
such that the scalar function $\mathbb{E}$ is dominant and $\mathbb{H}$ is practically negligible; the quark propagator is expressed as usual:
\begin{equation}
    \label{eq:quarkPropDef}
    S_f(p)=  Z_f(p^2) (i \gamma \cdot p + M_f(p^2))^{-1}\;;
\end{equation}
written in this way, one can establish the analogy with the tree level quark propagator, with $Z_f(p^2)$ and the mass function, $M_f(p^2)$, encoding the non-perturbative effects from the strong interactions of QCD (for instance, dynamical mass generation).

The BSWF entering Eq.~\eqref{eq:defLFWF} might be obtained following continuum schwinger methods (CSM) and, in particular, within the Dyson-Schwinger and Bethe-Salpeter equations formalism~\cite{Roberts:1994dr,Eichmann:2016yit}. This is a well-established approach in which predictions for $\pi$-$K$ electromagnetic form factors~\cite{Chang:2013nia,Raya:2015gva,Gao:2017mmp,Eichmann:2019bqf,Miramontes:2021exi}, as well as PDAs and PDFs~\cite{Chang:2013pq,Shi:2014uwa,Shi:2015esa,Ding:2019lwe,Ding:2019qlr,Cui:2020dlm,Cui:2020tdf,Cui:2021mom,Cui:2022bxn}, are readily available. Some explorations on LFWFs and GPDs are also taking place~\cite{Eichmann:2021vnj,Shi:2021nvg}. Herein, taking advantage of Eqs.~(\ref{eq:PDALFWF}, \ref{eq:uMzetaH}), we follow the discussion from Refs.~\cite{Raya:2021zrz,Zhang:2021mtn} to derive $\psi_{\textbf{P}u}^{\uparrow \downarrow}$ in a numerically accessible yet insightful way.

\subsection{Perturbation theory integral representation}
\begin{figure}[H]
 \centering
 \includegraphics[width=0.9\linewidth]{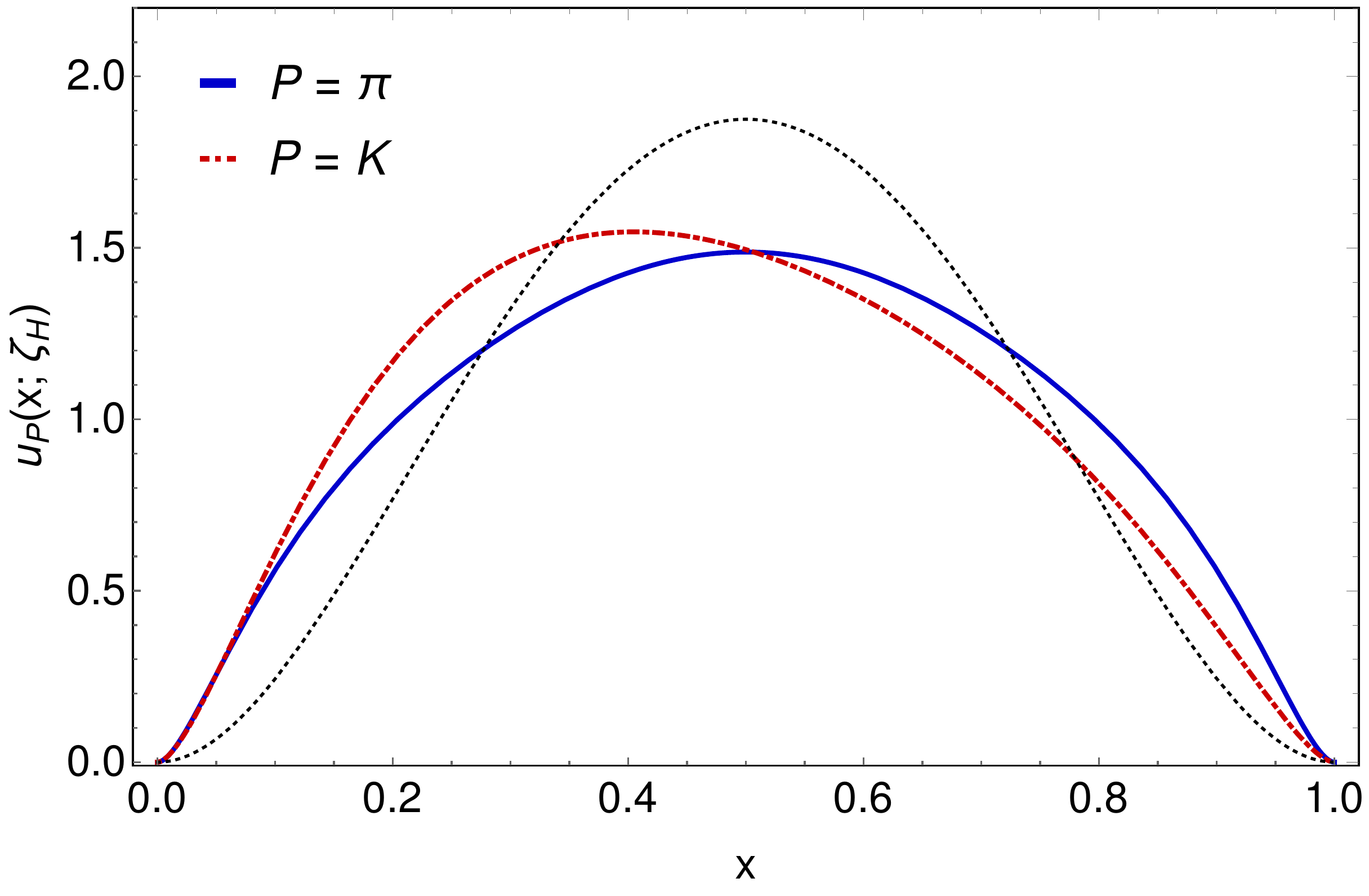}\\
 \caption{$\pi^+$ and $K^+$ valence-quark PDFs at $\zeta_H$. The PDFs are dilated with respect to $u^{sf}(x)$ (black dotted curve), Eq.~\eqref{eq:asymPDFs}, as a consequence of EHM. The skewness in $u_K(x;\zeta_H)$ is moduled by the $M_s - M_u$ difference, the Higgs interplay, producing $\textless x \textgreater_{K}^{u}$ = 0.47 and $\textless x  \textgreater_{K}^{s}$ = 0.53; in the pion case, the momentum distribution is equitable. }

 \label{fig:PDFs}
\end{figure}

Within our perturbation theory integral representation (PTIR) approach, the quark propagator and BSA read as: 
%
\begin{eqnarray}\label{eq:quarkparam}
S_f(k) & =&[ -i \gamma\cdot k + M_f] \Delta(k^2,M_f^2)\,,\\
{\mathpzc n}_{\textbf{P}} \Gamma_{\textbf{P}}(k) & =&  i \gamma_5  \int_{-1}^1\,dw\,\rho_{\textbf{P}}(w) \hat \Delta(k_w^2,\Lambda_{\textbf{P}}^2)\,,
\label{eq:BSAparam}
\end{eqnarray}
where
$M_f$ is interpreted as a constituent quark mass (the value at $p^2\simeq 0$ of the mass function, Eq.~\eqref{eq:quarkPropDef}) and $\Lambda_{\textbf{P}}$ is a mass-dimension parameter;
$\Delta(s,t) = 1/[s+t]$, $\hat \Delta(s,t) = t \Delta(s,t) $;
$k_w = k+ (w/2) P_{\textbf{P}}$, with $P^2_{\textbf{P}}=-m_{\textbf{P}}^2$, 
and $\rho_{\textbf{P}}(w)$ is a spectral weight whose particular profile determines the corresponding meson's BSA, with ${\mathpzc n}_{\textbf{P}}$ the related canonical normalisation constant. The Dirac structure in Eq.~\eqref{eq:BSAparam}, as compared with that in Eq.~\eqref{eq:BSAstruct}, implies that only the leading BSA is being retained~\footnote{Others can be incorporated systematically but, for practical purposes, all crucial information can be conveniently captured in the model parameters and spectral weight.}. Hence, the meson's BSWF can be readily recast~\cite{Xu:2018eii}:
\begin{subequations}
\begin{align}
{\mathpzc n}_\textbf{P} \chi_{\textbf{P}}\left(k_-;P\right) & =
{\mathpzc M}(k;P)
\int_{-1}^1\,dw\,\rho_{\textbf{P}}(w) {\mathpzc D}(k;P)\,,\\
\nonumber
{\mathpzc M}(k;P)  & = -\gamma_5[ \gamma\cdot P  M_{u}+ \gamma\cdot k (M_u-M_h) \\
& \qquad + \sigma_{\mu\nu} k_\mu P_{\nu}] \,,\\
\nonumber {\mathpzc D}(k;P) & =  \Delta(k^2,M_u^2) \Delta((k-P)^2,M_h^2) \\
 & \qquad \times \hat\Delta(k_{w-1}^2,\Lambda_{\textbf{P}}^2)\,;
\end{align}
\end{subequations}
After introducing two Feynman parameters ($\alpha$, $v$), the BSWF can be conveniently expressed as 
\begin{align}
\label{X2a}
\chi_{\textbf{P}}\left(k_-, P \right) & = {\mathpzc M}(k;P) \int_0^1 \,d\alpha\,2\, {\mathpzc X}_{\textbf{P}}(\alpha;\sigma^3(\alpha))\,,
\end{align}
with $\sigma(\alpha) = (k-\alpha P)^2+ \Omega_{\textbf{P}}^2$,
\begin{align}
\nonumber
\Omega_{\textbf{P}}^2 & = v M_u^2 + (1-v)\Lambda_{\textbf{P}}^2 \\
\nonumber
& + (M_h^2-M_u^2)\left(\alpha - \tfrac{1}{2}[1-w][1-v]\right) \\
&  + ( \alpha [\alpha-1] + \tfrac{1}{4} [1-v] [1-w^2]) m_{\textbf{P}}^2\,,
\label{Omega}\\
\nonumber
{\mathpzc X}_{\textbf{P}}(\alpha;\sigma^3) & =
\left[
\int_{-1}^{1-2\alpha} \! dw \int_{1+\frac{2\alpha}{w-1}}^1 \!dv \right. \\
&\quad  \left. + \int_{1-2\alpha}^1 \! dw \int_{\frac{w-1+2\alpha}{w+1}}^1 \!dv \right]\frac{\rho_{\textbf{P}}(w) }{{\mathpzc n}_{\textbf{P}} } \frac{\Lambda_{\textbf{P}}^2}{\sigma^3}\,.\label{X2c}
\end{align}
Appealing to Eq.\,\eqref{eq:defLFWF} and the Mellin moments of the distribution, 
\begin{equation}
    \langle x^m \rangle_{\psi_{\textbf{P}u}^{\uparrow \downarrow}}^{\zeta_H} = \int_0^1 dx\, x^m \psi_{\textbf{P}u}^{\uparrow \downarrow}(x,k_\perp^2;\zeta_H) \;,
\end{equation}
a series of algebraic manipulations yield:
\begin{eqnarray}
 \int_0^1 dx \, x^m\,  \psi_{\textbf{P}u}^{\uparrow\downarrow}(x,k_\perp^2;\zeta_H)
=  \int_0^1 d\alpha \, \alpha^m\, \mathcal{F}_\textbf{P}(\alpha){\mathpzc X}_{\textbf{P}}(\alpha;\sigma_\perp^2)\,; \nonumber 
\label{eq:xnLFWF}
\end{eqnarray} 
then, the uniqueness of Mellin moments enables us to relate the Feynman parameter $\alpha$ with the momentum fraction $x$, and identify:
\begin{eqnarray}
\label{eq:LFWFPTIR}
    \psi_{\textbf{P}u}^{\uparrow\downarrow}(x,k_\perp^2;\zeta_H) &=& \mathcal{F}_\textbf{P}(x){\mathpzc X}_{\textbf{P}}(x;\sigma_\perp^2)\;,\\
    \label{eq:Ftr}
     \mathcal{F}_{\textbf{P}}(x)&=&12[M_u+x(M_s-M_u)]\,.
\end{eqnarray}
From Eqs.~(\ref{eq:PDALFWF}, \ref{eq:uMzetaH}), it is now evident the impact of the spectral weight in the profiles of PDAs and PDFs. For instance, in the chiral limit ($m_\textbf{P}=0$,  $\Lambda_{\textbf{P}}=M_h=M_u$), the choice
\begin{equation}
    \rho^{as}(\omega) = \frac{3}{4}(1-\omega^2)\,
\end{equation}
yields the asymptotic distributions~\cite{Mezrag:2016hnp,Chouika:2017rzs,Chouika:2017dhe}:
\begin{eqnarray}
\label{eq:asymPDFs}
\varphi^{as}(x) = 6x(1-x) \;,\;u^{sf}(x) = 30 x^2(1-x)^2\;;
\end{eqnarray}
the latter also referred to as the scale-free parton-like profile PDF. If the most sophisticated CSM predictions currently available for $\pi-K$ PDAs and PDFs are employed to inform the choice of $\rho_\textbf{P}(\omega)$ (\emph{e.g.}, Refs.\,\cite{Chang:2013pq,Ding:2019lwe,Ding:2019qlr,Shi:2014uwa,Shi:2015esa,Cui:2020dlm,Cui:2020tdf}), then, the parametrization of the spectral weight introduced in\,\cite{Xu:2018eii} turns out to be suffiently flexible; this reads:
\begin{align}
n_{\rho} \rho_{\textbf{P}}(\omega) 
&=\frac{1}{2b_0^{\textbf{P}}}\left[\mbox{sech}^2 \left(\frac{\omega-\omega_0^{\textbf{P}}}{2b_0^{\textbf{P}}}\right)+\mbox{sech}^2 \left(\frac{\omega+\omega_0^{\textbf{P}}}{2b_0^{\textbf{P}}}\right)\right] \nonumber \\ 
&\times    (1+\omega\; v_{\textbf{P}})\;,
\label{eq:spectralw}
\end{align}
where $b_0^{\textbf{P}},$ $\omega_0^{\textbf{P}},\;v^{\textbf{P}}$ control the weight's profile (the latter, specially, sizes the $M_h-M_u$ flavor asymmetry), and $n_{\rho}$ is a derived normalization constant. The choice of parameters employed herein is found in~\cite{Raya:2021zrz}; our preferred values $M_u=0.31$ GeV and $M_s=1.2\, M_u$ are motivated by realistic solutions of the Dyson-Schwinger equation for the quark propagator~\cite{AtifSultan:2018end,Binosi:2016wcx}. The distributions in Fig.~\ref{fig:PDFs}, derived in Refs.~\cite{Cui:2020dlm,Cui:2020tdf}, are employed as benchmarks.

\begin{figure}[H]
 \centering
 \includegraphics[width=0.8\linewidth]{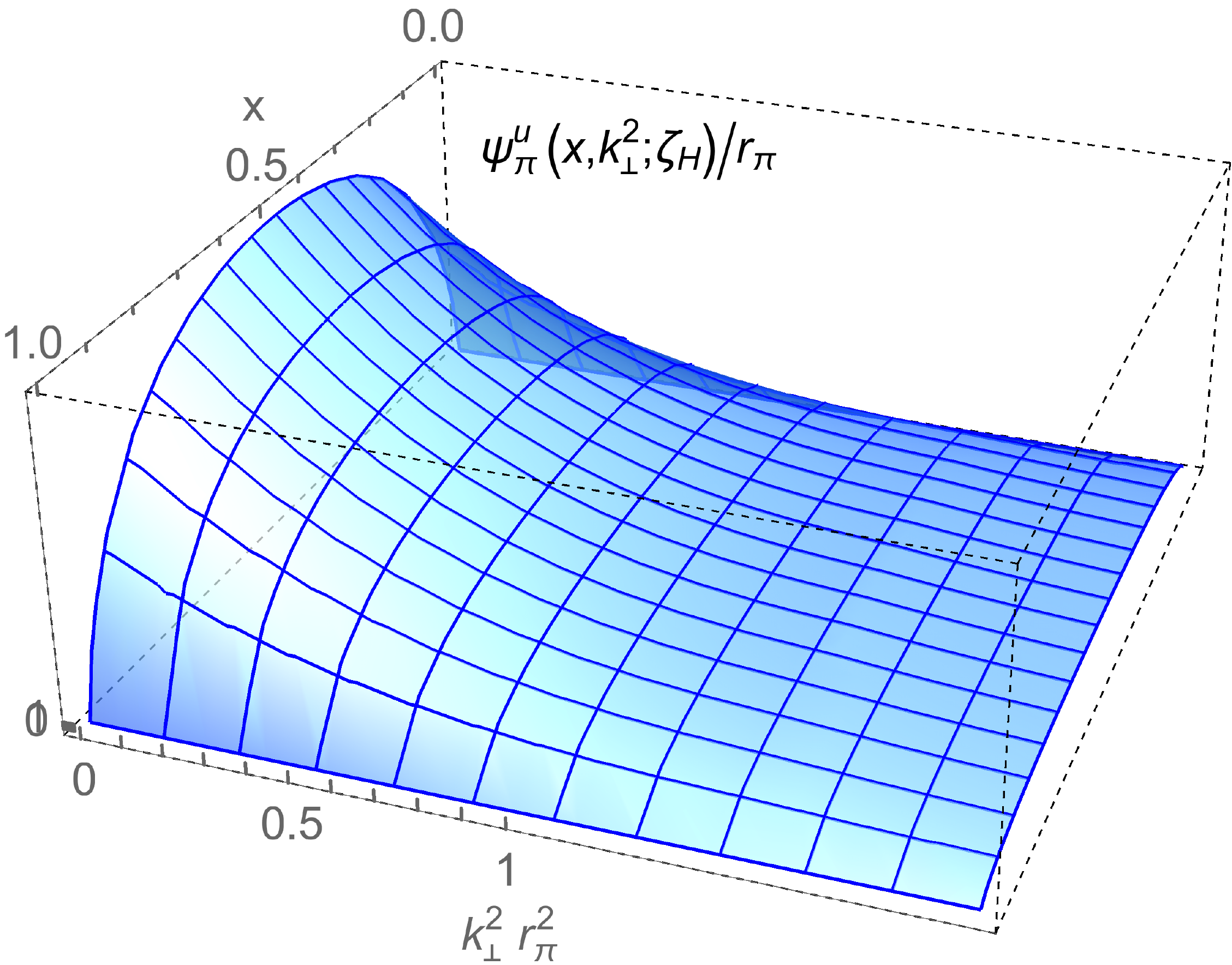}\\
 \includegraphics[width=0.8\linewidth]{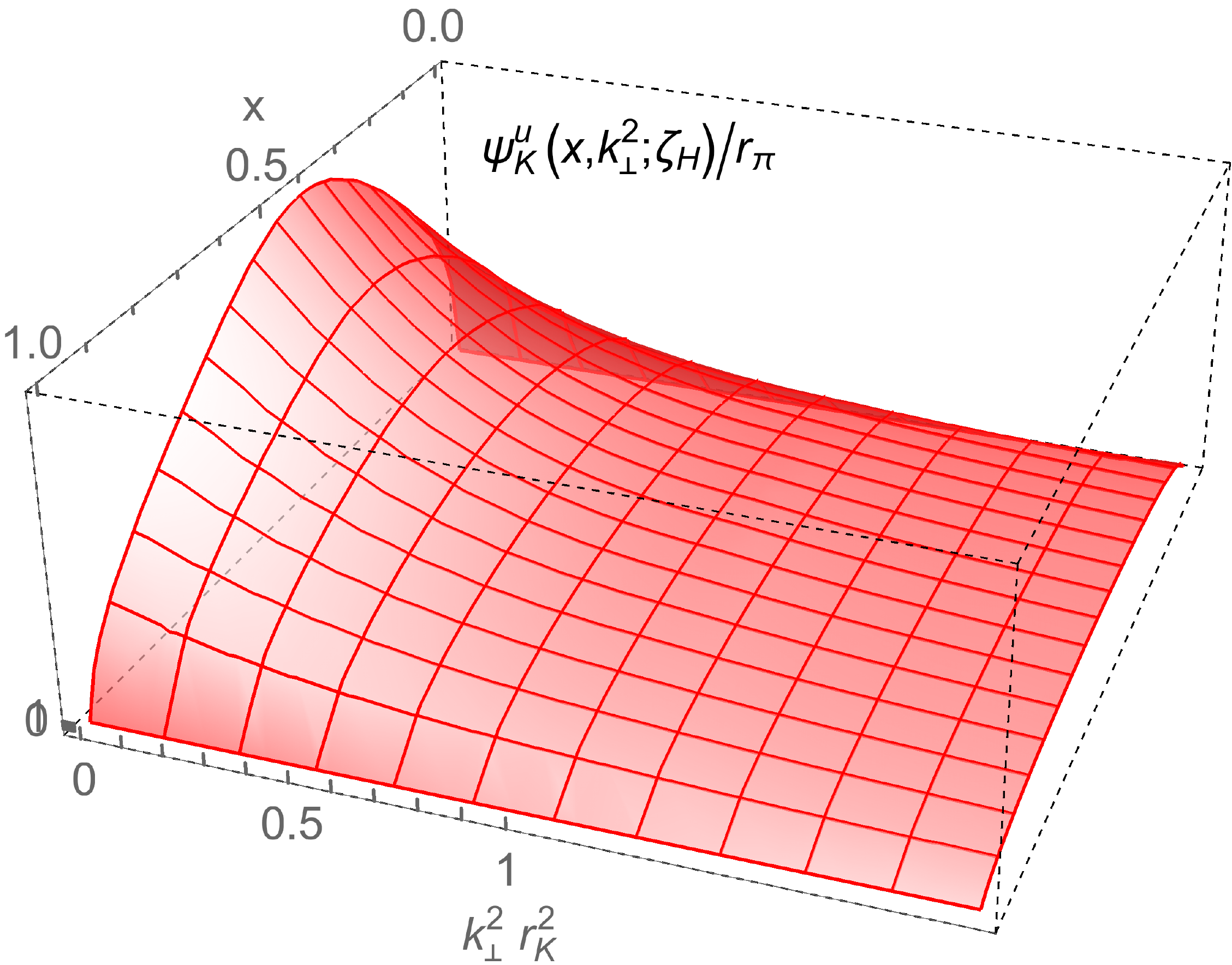}\\
 \caption{$\pi^+$ and $K^+$ leading-twist LFWFs in the PTIR approach, using the spectral weight defined in Eq.\eqref{eq:spectralw} and the parameter sets from Ref.~\cite{Raya:2021zrz}. }
 \label{fig:LFWFs}
\end{figure}

\begin{figure}[H]
 \centering
 \includegraphics[width=0.8\linewidth]{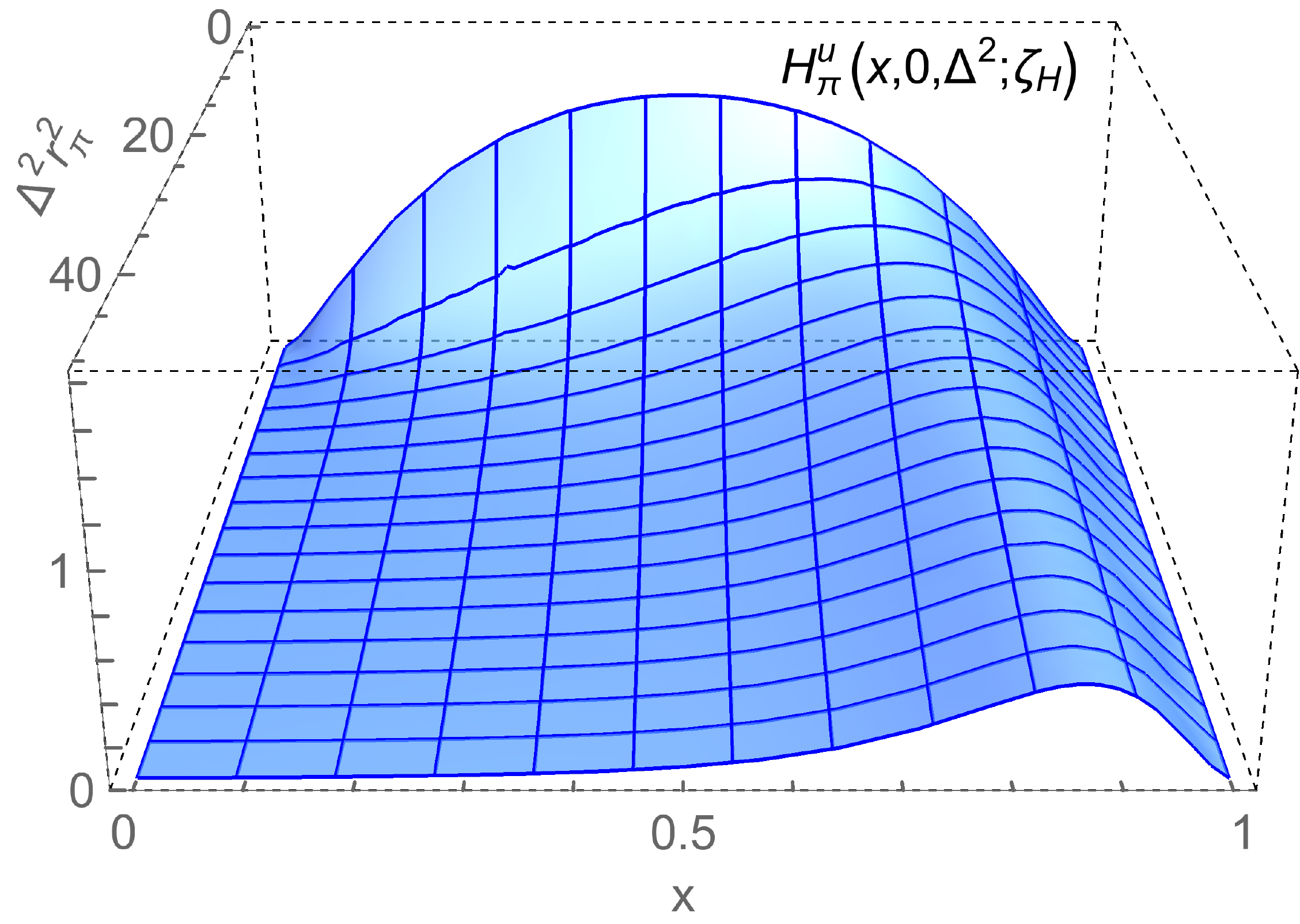}\\
 \includegraphics[width=0.8\linewidth]{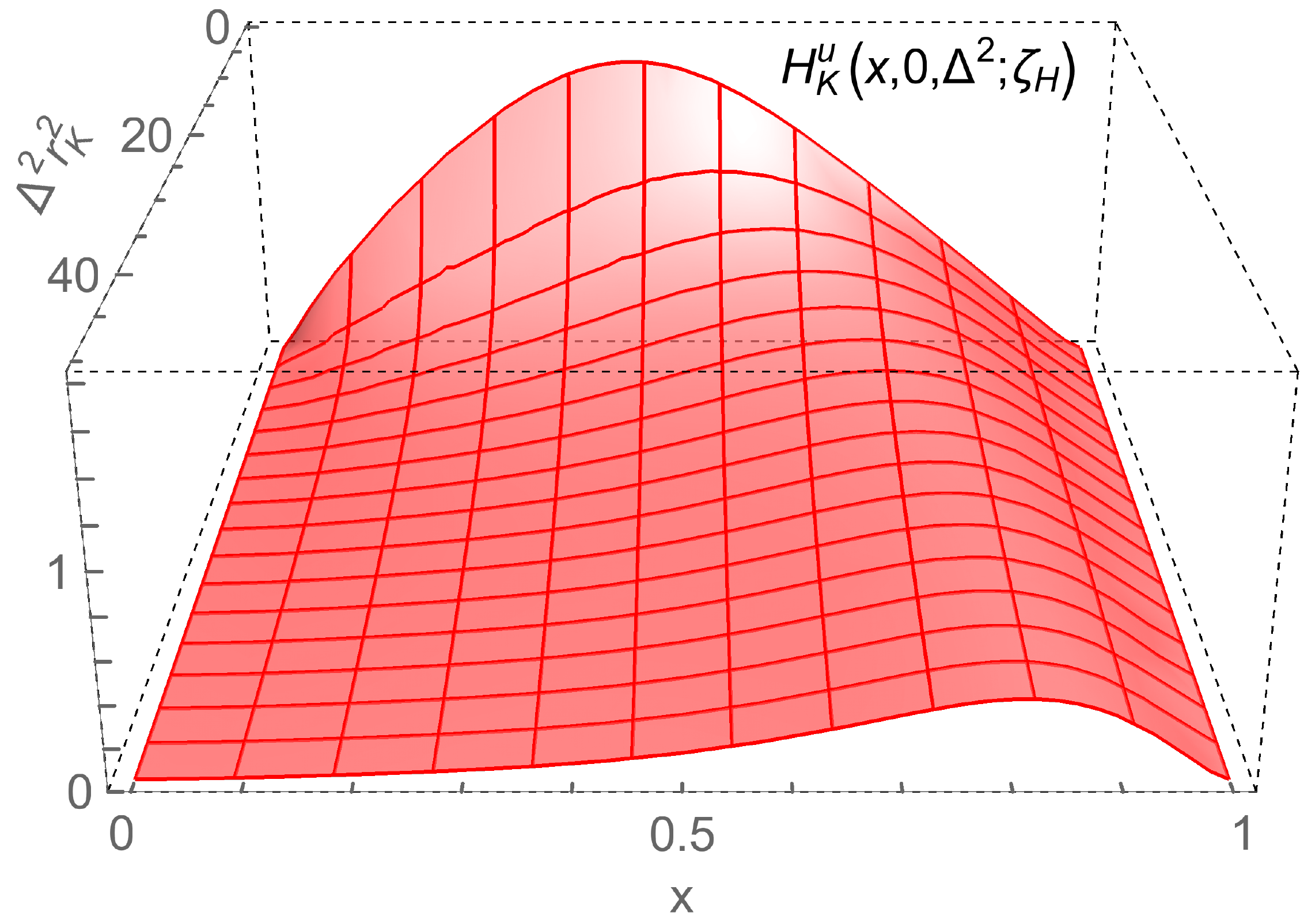}\\
 \caption{$\pi^+$ and $K^+$ valence-quark GPDs at $\zeta_H$, obtained from the overlap representation of the LFWFs, Eq.~\eqref{eq:overlap}. }
 \label{fig:GPDs}
\end{figure}
\section{Generalized parton distributions}
The valence-quark GPD ($u$-in-$\textbf{P}$) might be obtained from the leading-twist LFWFs by appealing to the so-called overlap representation~\cite{Diehl:2003ny}:
\begin{eqnarray}\nonumber
H^u_{\textbf{P}}(x,\xi,t;\zeta_H) = \hspace{5cm}
\\ \int \frac{d^2{\bf k_\perp}}{16 \pi^3} 
\psi^{\uparrow \downarrow \ast}_{\textbf{P}u}\left(x_-,\hat{\bf k}_\perp^2;\zeta_H \right)  
\psi^{\uparrow \downarrow}_{\textbf{P}u}\left( x_+,\tilde{\bf k}_\perp^2;\zeta_H \right) \,,
\label{eq:overlap}
\end{eqnarray} 
where $P$ ($\Delta$) is the momentum average (transfer) of the hadron states, $t=-\Delta^2$ and $\xi=-\Delta/[2 P^+]$ is the longitudinal momentum fraction transfer, usually dubbed skewness; as usual, $x$ is the longitudinal momentum fraction average and $\Delta_\perp^2=\Delta^2(1-\xi^2)-4\xi^2 m_\textbf{P}^2$. The boosted incoming ($-$) and outgoing ($+$) kinematic variables are defined as follows:
\begin{eqnarray}
&\displaystyle x_- = \frac{x-\xi}{1-\xi}\,, & \rule[0cm]{0.5cm}{0cm} \hat{\bf k}_\perp = {\bf k}_\perp + \frac{1-x}{1-\xi} \frac{{\bf \Delta}_\perp}{2} \;;\nonumber \\
&\displaystyle x_+ = \frac{x+\xi}{1+\xi} \,, & \rule[0cm]{0.5cm}{0cm} \tilde{\bf k}_\perp = {\bf k}_\perp - \frac{1-x}{1+\xi}
\frac{{\bf \Delta}_\perp}{2} \;; 
\label{eq:hatandtilde}
\end{eqnarray}
One can immediately realize that Eq.\,\eqref{eq:uMzetaH}, which makes the contact of the meson's PDF with its LFWF, results from specializing Eq.\,\eqref{eq:overlap} for the forward limit (${\bf \Delta}=0$, $\xi=0$), in which there is no boost and thus $x_-=x_+=x$ and $\hat{\bf k}_\perp=\tilde{\bf k}_\perp={\bf k}_\perp$.

It is worth stressing that the overlap representation is only valid within the DGLAP kinematic domain ($|x|\geq \xi$) and, additionally, the quark (antiquark) GPD is non-zero only for $x \geq -\xi$ ($x \leq \xi$).  A complete definition of the GPD within the overlap approach then requires an extension of Eq.\,\eqref{eq:overlap} to $|x| \leq \xi$, known as ERBL kinematic region. This is a very challenging task for which some progress has been recently seen\,\cite{Chouika:2017dhe,Chouika:2017rzs,Chavez:2021llq,Chavez:2021koz}. Furthermore, as time-reversal invariance guarantees for the so-defined GPD its being $\xi$-even, we will therefore restrict to $\xi \geq 0$ in the following. Our produced GPDs are displayed in Fig.~\ref{fig:GPDs}.
 
 \subsection{Electromagnetic and gravitational form factors}
\begin{figure}[H]
 \centering
 \includegraphics[width=0.8\linewidth]{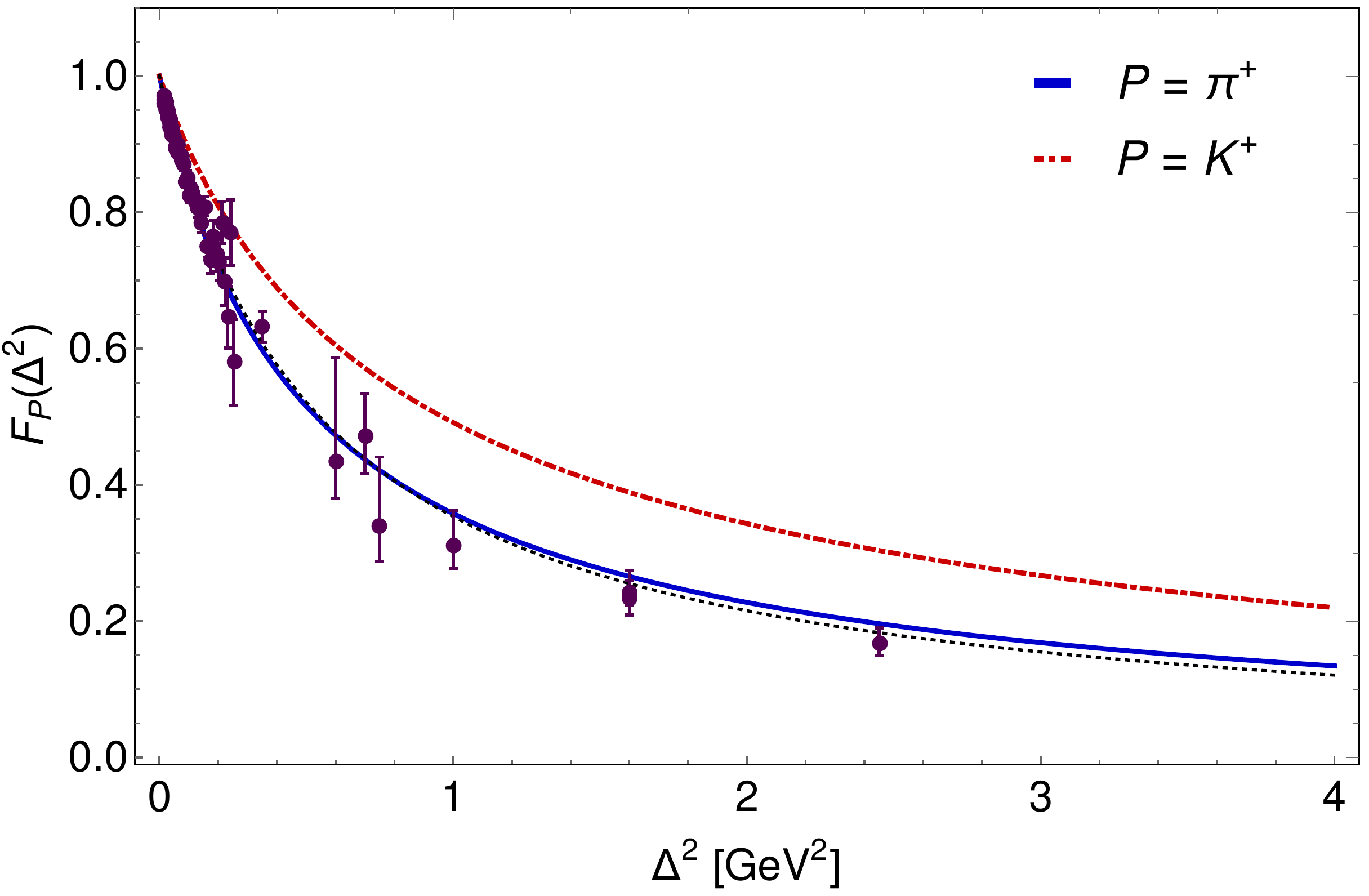}\\
 \caption{$\pi^+$ and $K^+$ EFFs, which produce $r_{\pi^+} = 0.69$ fm and $r_{K^+} = 0.56$ fm. The CSM result on $\pi^+$~\cite{Miramontes:2021exi} has been included for comparisson (black dotted line). Experimental data from Refs.~\cite{JeffersonLab:2008gyl,NA7:1986vav}.}
 \label{fig:EFFs}
\end{figure}

 The $u$-in-$\textbf{P}$ contribution to the meson's EFF is obtained from the zeroth moment of the GPD:
\begin{equation}
\label{eq:EFFdef1}
F_\textbf{P}^u(\Delta^2)=\int_{-1}^1 dx\, H_\textbf{P}^u(x,\xi,-\Delta^2;\zeta_H),
\end{equation}
The result is independent of $\xi$ and thus one can safely take $\xi=0$; furthermore, evolution equations entail that the zeroth moment is invariant under scale evolution. The complete meson EFF follows after summing over the valence contributions, weighed by the corresponding electric charges $e_{u,\bar{h}}$:
\begin{equation}
F_\textbf{P}=e_u F_\textbf{P}^u(\Delta^2)+e_{\bar{h}}F_\textbf{P}^h(\Delta^2)\;.
\end{equation}
The corresponding charge radius is defined as usual
\begin{equation}
\label{eq:chargeradii}
r_\textbf{P}^2 = -[6/F_\textbf{P}(0)]dF_\textbf{P}(\Delta^2)/d\Delta^2|_{\Delta^2=0}\;,
\end{equation}
such that, we obtain $r_{\pi^+} = 0.69$ fm and $r_{K^+} = 0.56$ fm, in fair agreement with the empirical estimates~\cite{ParticleDataGroup:2020ssz}. The produced EFFs are shown in Fig.~\ref{fig:EFFs}. 

Gravitational form factors (GFFs) are accessed via the first moment of the GPD:
\begin{equation}
\label{eq:GFFsdef1}
\int_{-1}^1dx x\, H_\textbf{P}^u(x,\xi,-\Delta^2;\zeta_H) = \theta_2^{\textbf{P}u}(\Delta^2)-\xi^2 \theta_1^{\textbf{P}u}(\Delta^2)\;,
\end{equation}
where $\theta_1$ is related to pressure and shear forces inside the hadron, and $\theta_2$ to the mass distribution~\cite{Polyakov:2018zvc}. The individual form factors on the right-hand-side of Eq.~\eqref{eq:GFFsdef1} are scale invariant, but the left-hand side that changes under QCD evolution, exposing contributions coming from sea-quarks and gluons~\cite{Raya:2021zrz}. The complete meson GFFs is obtained by summing up the independent contributions.  Note that, while $\theta_2$ is readily accesible by taking $\xi=0$, whereas $\theta_1$ demands the knowledge of the GPD in the ERBL region; a particular way to overcome such problem has been introduced in~\cite{Zhang:2020ecj}. 

For the sake of brevity, and because their profiles are very similar to that of the EFFs, GFFs are not displayed herein ( the reader is referred to Sec. VI of~\cite{Raya:2021zrz} instead). Nonetheless, it is important to highlight the ordering of hardness of the form factors, as inferred from the corresponding radii (obtained in analogy with  Eq.~\eqref{eq:chargeradii}):
\begin{equation}
\label{eq:ordering}
    r_\textbf{P}^{\theta_2} \approx 0.81\, r_\textbf{P} \; \textless \; r_\textbf{P}\; \textless \; r_\textbf{P}^{\theta_1 } \approx 1.18\, r_\textbf{P}\;.
\end{equation}
 The above reveals that charge effects span over a larger domain than mass effects, which is as also manifested in the charge and mass distributions discussed below~\footnote{The first inequality can be proven algebraically for the so called factorized LFWFs.}. It is also observed that the radii associated to $K^+$ are compressed by $\sim 15\%$ with respect to $\pi^+$.

\begin{figure}[H]
 \centering
 \includegraphics[width=0.8\linewidth]{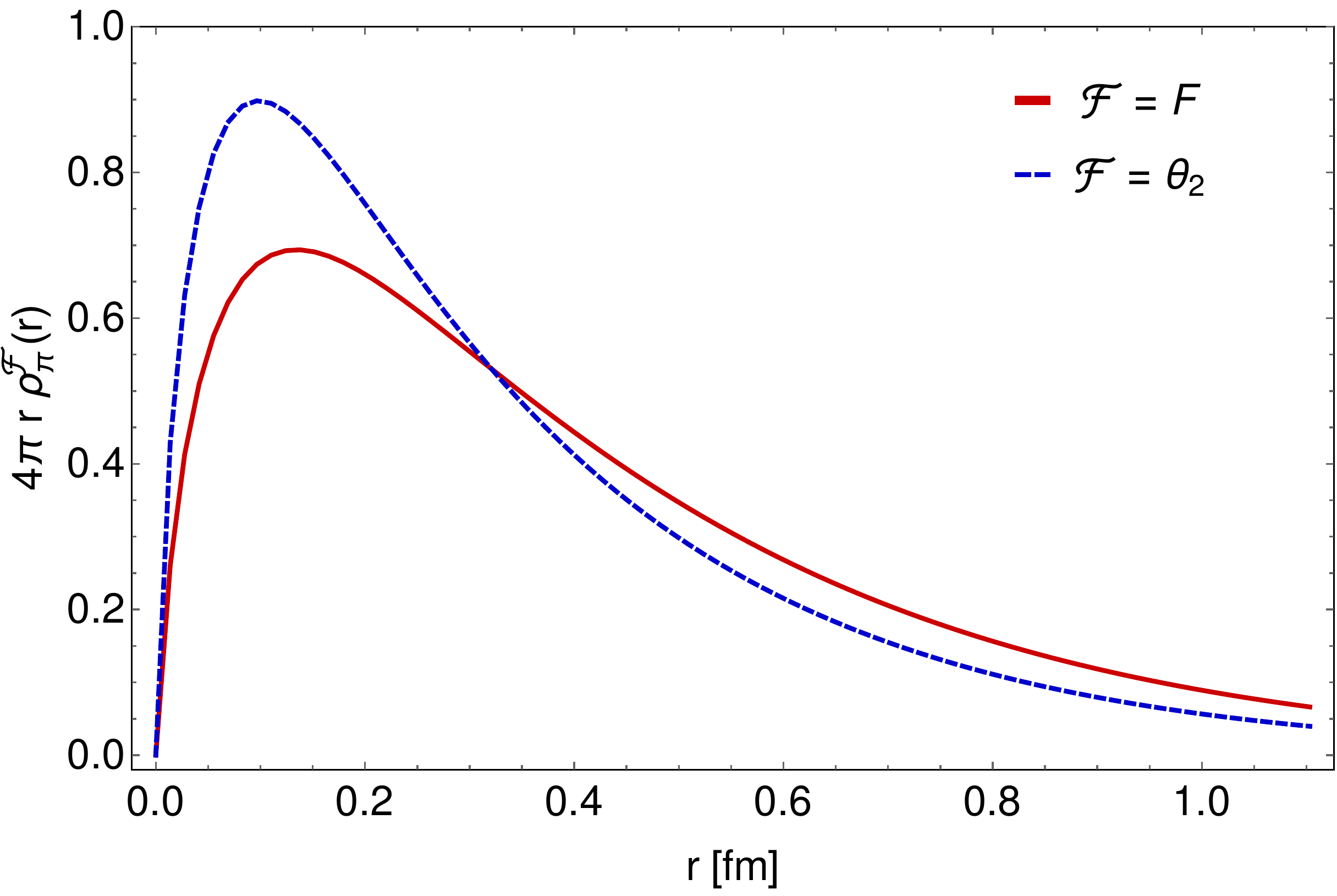}\\\hspace{0.2cm}
 \includegraphics[width=0.85\linewidth]{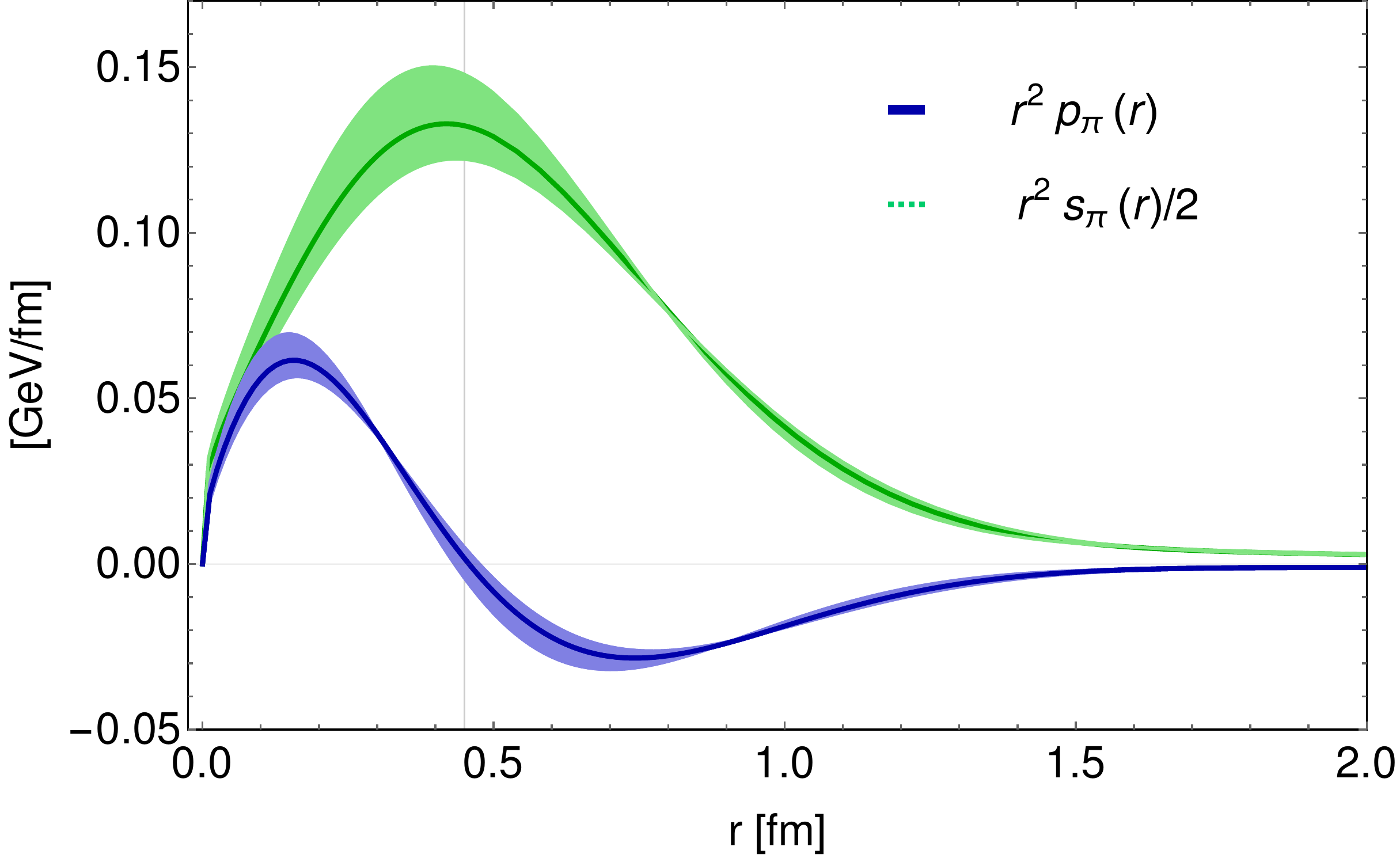}\\
 \caption{Upper panel: $\pi^+$ charge and mass distributions, confirming the intuition given by the inequality $r_\textbf{P}^{\theta_2} \textless r_\textbf{P}$, Eq.~\eqref{eq:ordering}. Lower panel: pressure and shear distributions, Eq.~\eqref{EqPressureB}. The pressure changes sign around $r_c=0.45$ fm, indicating the domain in which confinement effects become dominant; shear forces are maximal in the neighbourhood. $K^+$ results (not displayed herein) exhibit similar patterns, albeit the distributions are more compressed.}
 \label{fig:CMPdists}
\end{figure}

\subsection{Charge, mass and pressure distributions}
Images of the charge and mass distributions within the mesons might be obtained by Fourier transforming the corresponding form factors, $F_\textbf{P}(\Delta^2)$ and $\theta_2^{\textbf{P}}(\Delta^2)$, as follows:
\begin{equation}
    \label{eq:chargeandmass}
    \rho_\textbf{P}^{\mathfrak{F}}(r)=\frac{1}{2\pi} \int_0^\infty d\Delta \Delta J_0(\Delta r) \mathfrak{F}_\textbf{P}(\Delta^2)\;,
\end{equation}
where $J_0$ is a Bessel function of the first kind and $\mathfrak{F}_\textbf{P}=F_\textbf{P},\theta_2^\textbf{P}$. Pressure and shear profiles ($p_\textbf{P}(r)\,\text{and}\,s_\textbf{P}(r)$, respectively) are derived from $\theta_1$ as follows:
\begin{subequations}
\label{EqPressure}
\begin{align}
p_\textbf{P}(r)  & =
 \frac{1}{6\pi^2 r} \int_0^\infty d\Delta \,\frac{\Delta}{2 E(\Delta)} \, \sin(\Delta r) [\Delta^2\theta_1^{\textbf{P}}(\Delta^2)] \,, \label{EqPressureA}\\
 s_\textbf{P} (r)  & =
%
\frac{3}{8 \pi^2} \int_0^\infty d\Delta \,\frac{\Delta^2}{2 E(\Delta)} \, {\mathpzc j}_2(\Delta r) \, [\Delta^2\theta_1^{\textbf{P}}(\Delta^2)] \,, \label{EqPressureB}
\end{align}
\end{subequations}
where $2E(\Delta)=\sqrt{4 m_{\textbf{P}}^2+\Delta^2}$ and ${\mathpzc j}_2(z)$ is a spherical Bessel function. Fig.~\ref{fig:CMPdists} displays the charge, mass and pressure distributions, associated to the electromagnetic and gravitational form factors.
\begin{figure}[H]
 \centering
 \includegraphics[width=0.8\linewidth]{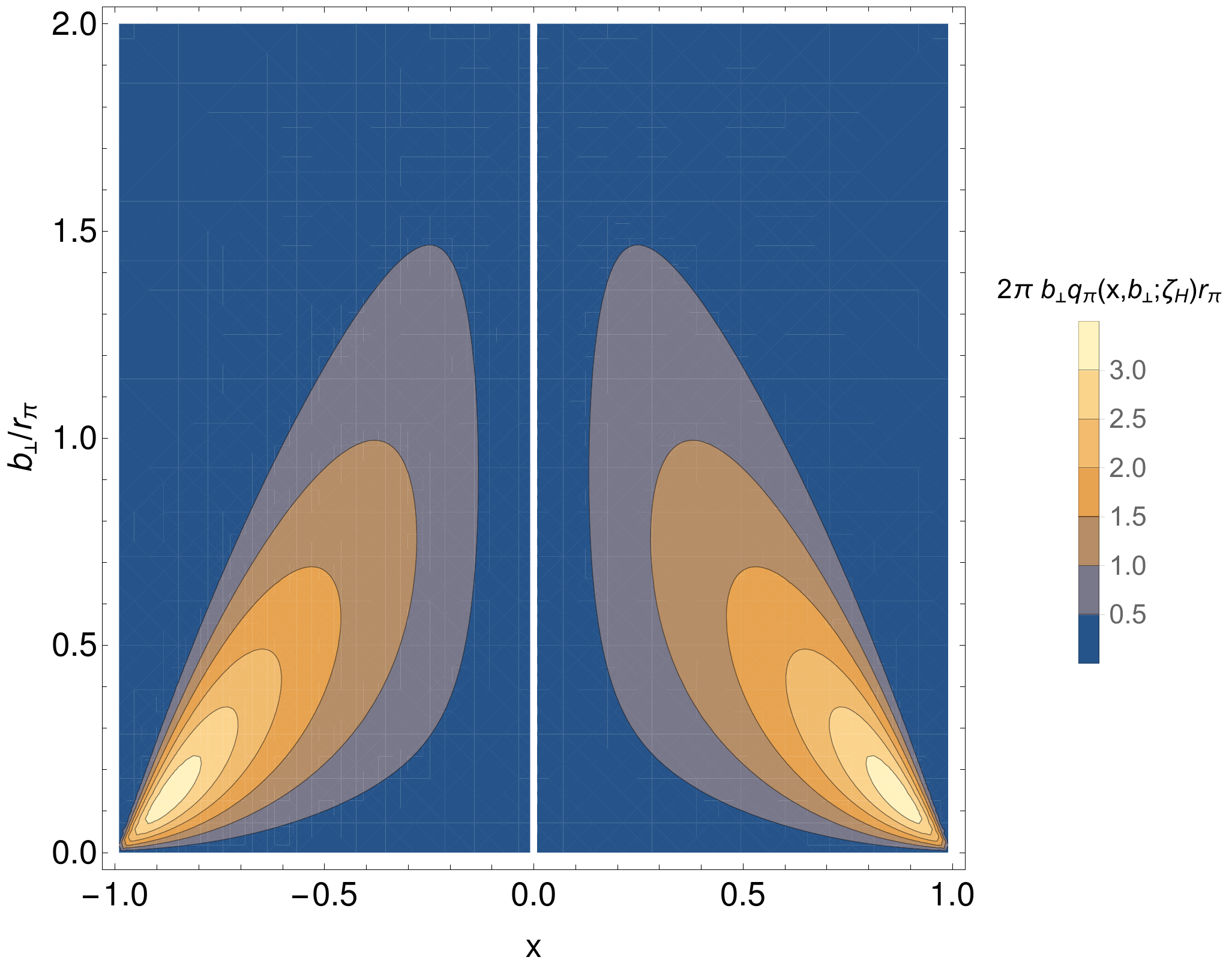}\\\hspace{0.2cm}
 \includegraphics[width=0.85\linewidth]{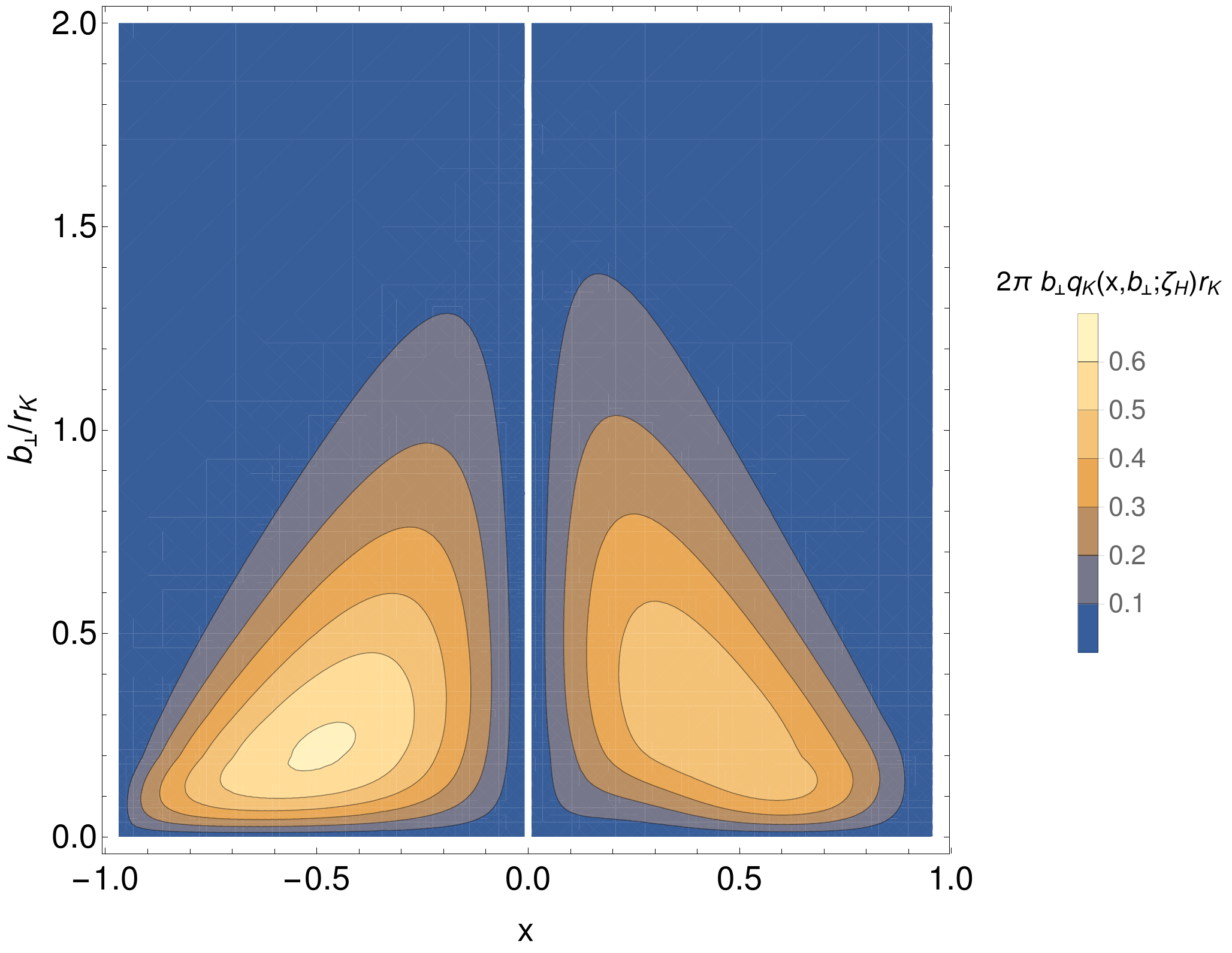}\\
 \caption{$\pi^+$ and $K^+$ IPS-GPDs at $\zeta_H$, Eq.~\eqref{eq:IPDHgen}. For the pion, the maximum is located at $(|x|,b_\perp/r_\pi)=(0.88,0.13)$; for the kaon, at $(x,b_\perp/r_K)_{\bar{s}}=(-0.87,0.13)$ and $(x,b_\perp/r_K)=(0.84,0.17)$. As the fully dress valence constituents become less dressed, via QCD evolution, the distributions flatten and the maximum shift towards $x\to 0$~\cite{Raya:2021zrz}; \emph{i.e.} they contribute less to the total momentum.}
 \label{fig:IPSGPDs}
\end{figure}

\subsection{Impact parameter space distributions}
The GPD can be defined in impact parameter space (IPS) as the Fourier transform in transverse momentum of the zero-skewness GPD~\cite{Burkardt:2002hr}, 
\begin{eqnarray}\nonumber
\mathpzc{u}^{\textbf{P}}(x,b_\perp^2;\zeta_H) &=&  \int_0^\infty \frac{d\Delta_\perp}{2\pi} \Delta_\perp J_0(b_\perp \Delta_\perp) \\ 
&\times& H_{\textbf{P}}^u(x,0,-\Delta_\perp^2;\zeta_H) \,.
\label{eq:IPDHgen}
\end{eqnarray} 
The IPS-GPDs, displayed in Fig.~\ref{fig:IPSGPDs}, describe the likelihood of finding a parton with a certain momentum fraction $x$, at a given transverse position $b_\perp$, defined with respect to the meson's center of transverse momentum (CoTM). The mean-squared transverse extent follows after the integration
\begin{eqnarray}
\langle b_\perp^2(x;\zeta_H)\rangle_\textbf{P}^u &=& \int d^2 b_\perp b^2_\perp \mathpzc{u}^{\textbf{P}}(x,b_\perp^2;\zeta_H) \\
&=& -4 \frac{\partial}{\partial \Delta_\perp^2} H_\textbf{P}^u(x,0,-\Delta_\perp^2;\zeta_H)|_{\Delta_\perp^2 = 0}\;, \nonumber
\end{eqnarray}
which, in the $\pi^+$ case  ($M_u = M_h$ in general), entails:
\begin{equation}
    \langle b_\perp^2(\zeta_H) \rangle_\textbf{P}^u = \int_0^1 dx \, \langle b_\perp^2(x;\zeta_H)\rangle_\textbf{P}^u  = \frac{2}{3}r_\textbf{P}^2\;;
\end{equation}
for the $K^+$, one obtains:
\begin{equation}
    \langle b_\perp^2(\zeta_H) \rangle_{K^+}^u = 0.69\, r_{K^+}^2\,,\, \langle b_\perp^2(\zeta_H) \rangle_{K^+}^s = 0.51 \,r_{K^+}^2\;,
\end{equation}
implying, naturally, that the $s$-in-$K$ plays a larger role in determining the CoTM.

\section{Insights from factorized LFWFs}
Even before specifying the spectral weight in Eq.~\eqref{eq:LFWFPTIR}, the form of $\mathcal{F}_\textbf{P}(x)$, Eq.~\eqref{eq:Ftr}, predicts an asymmetry in the kaon LFWFs and distributions derived therefrom. A further examination of $\Omega_{\textbf{P}}^2$ in Eq.\,\eqref{Omega} supports this observation, while also revealing $x-k_\perp$ correlations, proportional to $(M_h^2-M_u^2)$ and $m_\textbf{P}^2$. Consequently, such correlations vanish in the chiral limit and, in combination with Eqs.~(\ref{eq:PDALFWF}, \ref{eq:uMzetaH}), the LFWF can be recast as:
\begin{eqnarray}\nonumber
\psi_{\textbf{P}u}^{\uparrow\downarrow}(x,k_\perp^2;\zeta_H) &=&\tilde{\psi}_{\textbf{P}u}^\varphi (k_\perp^2) \varphi_\textbf{P}^u(x;\zeta_H) \\ &=& \tilde{\psi}_{\textbf{P}u}^\textbf{u} (k_\perp^2)\,[u^\textbf{P}(x;\zeta_H)]^{1/2}\;.\label{eq:facLFWF}
\end{eqnarray}
By corollary, one obtains the relation~\footnote{For $\zeta \geq \zeta_H$ this is no longer true, since PDA and PDF obey different evolution equations.}:
\begin{eqnarray}
\label{eq:PDFePDA2}
u^\textbf{P}(x;\zeta_H) = [\varphi_\textbf{P}^u(x;\zeta_H)]^{2}/\int_0^1 dx [\varphi_\textbf{P}^u(x;\zeta_H)]^{2} \;.
\end{eqnarray}
Eq.~\eqref{eq:facLFWF} define the so-called factorized LFWFs: the $k_\perp$ dependence is controlled by the profile functions $\tilde{\psi}_{\textbf{P}u}(k_\perp^2)$, while the $x$ dependence is driven, equivalently, by the PDA or PDF. Eq.~\eqref{eq:overlap} is then recast as:
\begin{equation}
H^u_{\textbf{P}}(x,\xi,t;\zeta_H) =\Theta(x_-) \sqrt{u^{\textbf{P}}(x_-;\zeta_H) u^{\textbf{P}}(x_+;\zeta_H)} \Phi_{\textbf{P}}(z;\zeta_H) \;,\label{eq:GPDfac}
\end{equation}
where $\Theta$ is the Heaviside funcation, $z=s_\perp^2= -t(1-x)^2/(1-\xi^2)^2$ and
\begin{equation}\label{eq:defPhi}
\Phi_{\textbf{P}}^u\left(z; \zeta_H \right) =
\int \frac{d^2{\bf k_\perp}}{16 \pi^3} 
 \widetilde{\psi}^{u \ast}_{\textbf{P}}\left({\bf k}_\perp^2;\zeta_H \right) 
\widetilde{\psi}^{u}_{\textbf{P}}\left(\left({\bf k}_\perp - \bf{s}_\perp \right)^2;\zeta_H \right)\;.
\end{equation}
The factorized LFWF, Eq.~\eqref{eq:facLFWF}, being obtained as the chiral limit of the PTIR one, Eq.~\eqref{eq:LFWFPTIR}, enables an algebraic evaluation of $\Phi_{\textbf{P}}^u$. Nonetheless, the combination of Eqs.~(\ref{eq:EFFdef1}, \ref{eq:GPDfac}, \ref{eq:defPhi}) yields the more general result:
\begin{eqnarray}\label{eq:PhifromF}
\langle x^{2n} \rangle_{\bar{h}}^{\zeta_H} \left. \frac{\partial^n}{\partial^n z} \Phi_{\textbf{P}}^u(z;\zeta_H) \right|_{z=0} &=& 
\left. \frac{d^n F_{\textbf{P}}^u(\Delta^2)}{d(\Delta^2)^n}  \right|_{\Delta^2=0} \;.
\end{eqnarray}
Then, when approaching the LFWF by a factorized \emph{Ansatz}, the $k_\perp^2$-dependence of the $u$-quark overlap GPD is plainly determined by the quark DFs and the EFF. An immediate outcome is the isospin symmetry limit relation:
\begin{equation}
    \left(\frac{r_\textbf{P}^{\theta_2}}{r_\textbf{P}}\right)^2 = \frac{\textless x(1-x) \textgreater_\textbf{P}^{\zeta_H}}{\textless x^2 \textgreater_\textbf{P}^{\zeta_H}}\,,
\end{equation}
which confirms the inequality of Eq.~\eqref{eq:ordering}, since the ratio of those moments is always positive and less than unity. More algebraic relations, specially for the IPS-GPDs and related distributions, are found in~\cite{Raya:2021zrz}.

\section{Conclusions and Scope}
Informed by the prior knowledge of 1-dimensional distributions (PDAs or PDFs), we have described an insightful approach to address several structural properties of pseudoscalar mesons via LFWFs and GPDs. The discussion focuses on pions and kaons, and their connection with the EHM. The latter manifests in PDAs and PDFs (as well as LFWFs and GPDs), broadened with respect to their asymptotic profiles. In turn, the interference with Higgs mass generation produces slightly skewed kaon distributions yielding a different arrangement in the mass/momentum distribution. Spatially, the kaon turns out to be slightly more compressed than the pion, as exposed by charge, mass, and pressure distributions (and corresponding radii). The role of QCD evolution has been omitted in the present discussion, but it is detailed in~\cite{Raya:2021zrz}. Finally, the algebraic approach that we have described herein can be systematically improved by completing both the BSA and LFWF descriptions; and extended in order to study other types of distributions (such as transverse momentum dependent distributions) and hadrons~\cite{draftMel}.

\section{Acknowledgments}
The authors acknowledge the kind invitation from the organizers of `The 19th International Conference on Hadron Spectroscopy and Structure in memoriam Simon Eidelman' (HADRON 2021), with whom they feel very grateful. Work supported by Spanish Ministry of Science and Innovation (MICINN) (grant no. PID2019-107844GB-C22) and Junta de Andaluc\'ia (grant nos. P18-FR-5057, UHU-1264517, UHU EPIT-2021).

\end{multicols}
\medline
\begin{multicols}{2}
\bibliography{main}
\bibliographystyle{unsrt}
\end{multicols}
\end{document}